# Damage identification of offshore jacket platforms in a digital twin framework considering optimal sensor placement

Mengmeng Wang[a, b], Atilla Incecik[b], Shizhe Feng[c], M.K. Gupta[d], Grzegorz Królczyk[d], Z Li[d, *]

a. Department of Marine Engineering, Ocean University of China, Qingdao 266100, China

b. Department of Naval Architecture, Ocean, and Marine Engineering, University of Strathclyde, Glasgow G11XQ, United Kingdom

c. School of Mechanical Engineering, Hebei University of Technology, Tianjin 300130, China

d. Faculty of Mechanical Engineering, Opole University of Technology, Opole, 45758, Poland

*Corresponding author: z.li@po.edu.pl

## Abstract

A new digital twin (DT) framework with optimal sensor placement (OSP) is proposed to accurately calculate the modal responses and identify the damage ratios of the offshore jacket platforms. The proposed damage identification framework consists of two models (namely one OSP model and one damage identification model). The OSP model adopts the multi-objective Lichtenberg algorithm (MOLA) to perform the sensor number/location optimization to make a good balance between the sensor cost and the modal calculation accuracy. In the damage identification model, the Markov Chain Monte Carlo (MCMC)-Bayesian method is developed to calculate the structural damage ratios based on the modal information obtained from the sensory measurements, where the uncertainties of the structural parameters are quantified. The proposed method is validated using an offshore jacket platform, and the analysis results demonstrate efficient identification of the structural damage location and severity.

## 1. Introduction

With the increasing dependence on renewable energy, the development of offshore wind energy has grown rapidly, evidenced by new installation capacity of 21.1GW in 2021, which is more than triple that in 2020 [1]. The rapid booming of offshore wind energy brings significant safety problems on offshore wind structures. This is because the offshore wind structures are directly exposed to harsh ocean environments [2], where corrosions and fatigue loads will cause frequent structural damages [3,4]. To protect the offshore wind structures, it is necessary to monitor their health conditions in real-time [5] and identify the damage at the early stage. Recently, the Digital twin (DT) technique [6–8] provides a promising solution for structural health monitoring (SHM). The DT is able to mirror model (called a DT virtual model) to simulate the dynamics of the physical unit in real-time [9–11] to implement the SHM.

A DT-based SHM framework generally includes four modules, i.e., one structure unit, one DT virtual model, one DT database and one SHM model. The structure unit serves as the basis for constructing the DT virtual model; the virtual model conducts simulations and analyses to collect the structure dynamic responses; the database stores the sensory measurements as well as the virtual model outputs to enable the dual information exchange between the structure unit and the virtual model; the SHM model monitors the structural health condition in real time, detects damages and issues timely alerts for maintenance and repairs. The DT-based SHM framework provides a comprehensive SHM tool for offshore wind structures. However, the development of such a framework is in an infant stage. Existing efforts have been made to build the DT virtual model, while very limited work has been done to address the SHM model in the DT framework [5]. The challenge here is that the modal information of the offshore wind structures in existing DT framework is assumed correct and effective, ignoring the model uncertainties; however, the model uncertainties of the structures always play an important role for precise damage identification. As a result, it is crucial to solving this issue when developing the DT-based SHM framework.

Since 1970s, the damage identification has been developed and many researchers have attempted different methods in damage detection and damage identification (the damage location and severity) of the offshore jacket platforms, such as the cross-model cross-mode (CMCM) method, machine learning [12–15], modal strain energy method and so on. However, most of the damage identification methods belong to deterministic methods, which cannot quantify the uncertainties of damage detection. To mitigate the negative effects of uncertainties on damage identification, probabilistic methods are often employed to quantify the impact of these uncertainties on accuracy [16]. Bayesian is a widely used probabilistic method in the literature and has been utilized for model updating, modal identification, and damage detection [17,18]. Bayesian analysis was combined with modal data by Huang et al. [19] to identify weakening of structural components' stiffness. In [20], Yin et al. introduced a numerical and experimental Bayesian framework for damage identification in the 2D frame. In order to identify damage severity using mode shapes in the simple beam element, Huang et al. [21] numerically implemented the Bayesian method. The effects of modelling errors on damage detection findings were studied by Behmanesh et al. [22], who used the Bayesian model updating technique to detect shear frame damage. Zhang et al. [23] used the Bayesian model updating technique, which takes into account model uncertainties, to detect and identify damage to the eight-story building. The Bayesian approach has been shown to be effective in addressing uncertainties in damage identification; however, to the best of our knowledge, very little work has been done to apply this approach to the problem of identifying damage to an offshore jacket platform. Since the structure of the offshore jacket platform is complicated, it is very challenging to determine the most important nodes that should be monitoring [24]. A sensor-optimization strategy must be applied to the offshore jacket platform when performing the Bayesian-based SHM. However, little work considers yet the optimal sensor placement (OSP) [32-34] in the Bayesian-based SHM framework.

The OSP problem is a min-max multi-objective optimisation problem [25,26] with the objectives of maximising some optimisation criteria while minimising the expense of sensor deployment [27]. The damage identification often depends on changes in modal information such as the natural frequency and mode shapes caused by the damage in the structural elements. Accurate acquisition of structural modal responses is critical for the damage identification and heavily depends on the OSP. However, the sensor installation in the offshore jacket platform is usually decided by experience; and hence, the OSP is not considered in most of the damage identification of the offshore jacket platforms [28–30], which will caused the waste of the sensors or the reduction of the accuracy of the measured modal information.

Generally, the OSP involves two aspects, the optimization criteria and optimization algorithms. In recent years, several optimization criteria have been developed [28,30–32]. Some aim to maximize the measuring merits, such as the Effective Independence (EI) [33], Kinetic Energy (KE) [34] and the Eigenvalue Vector Product (EVP) [35]; some other criteria aim to optimize the quality of the modal identification, such as the modal assurance criterion [36], mutual information [37] and information entropy (IE) [38,39]. However, when applying these criteria to the offshore jacket platforms, the number of sensors is fixed in the OSP procedure; as a result, one cannot explain the influence of the sensor number on each criterion, but also cannot balance the sensor costs and the OSP quality. It is always difficult to make a good trade-off between reducing the sensor number and maintaining high OSP quality. Hence, it is impossible to fix the sensor number during the OSP procedure for the offshore jacket platforms.

The optimisation method determines the OSP's effectiveness and reliability. The population strategy can be used to classify the many algorithms suggested to search the optimal sensor configuration into two groups (for example, the Genetic algorithm (GA) [40], Ant colony optimization [41], Particle swarm optimization (PSO) [42], Wolf algorithm [43] and Non-Dominated Sorting Genetic Algorithm II (NSGA-II) [44]) and the trajectory strategy (e.g. the Simulated Annealing). Recently, the multi-objective Lichtenberg algorithm (MOLA) is proposed [45,46] for the first time to integrate the trajectory and population strategies, which improves its convergence and maximum spread in the multi-objective optimization, and can generate Pareto optimal solutions considering multiple objectives. Moreover, the MOLA has demonstrated better convergence and coverage in the complex optimization problems in the CEC 2009 test functions and ZDT test functions than existing optimization algorithms [45]. It is worth investigating the MOLA as the optimization algorithm in the OSP process for the offshore jacket platforms. The quality of the OSP determines the accuracy of the collected modal information and the damage identification at the same time. Therefore, the OSP must be considered in the damage identification process.

To solve these problems, this paper proposes the first comprehensive methodology to identify damage to offshore jacket platforms while also taking into account the OSP. To strike a good balance between sensor cost and modal information accuracy, the OSP model takes into account the number of sensors and one of four well-known modal criteria (EFI, KE, EVP, and IE) as the optimised objectives to determine the sensor configurations (number and locations). Bayesian analysis is used in the damage identification model to compute the damage ratio from the modal data and to quantify the uncertainties of the model parameters. The proposed technique for damage identification of offshore jacket platforms shows promise in numerical evaluation findings.

This work is organized as follows. Section 2 provides a detailed description of the proposed damage identification framework considering the optimal sensor placement. The effectiveness of the proposed OSP method is validated in Section 3. Section 4 brings the results and discussions of damage identification, and Section 5 draws the main conclusions.

## 2. Methodology

### 2.1 Proposed damage identification framework

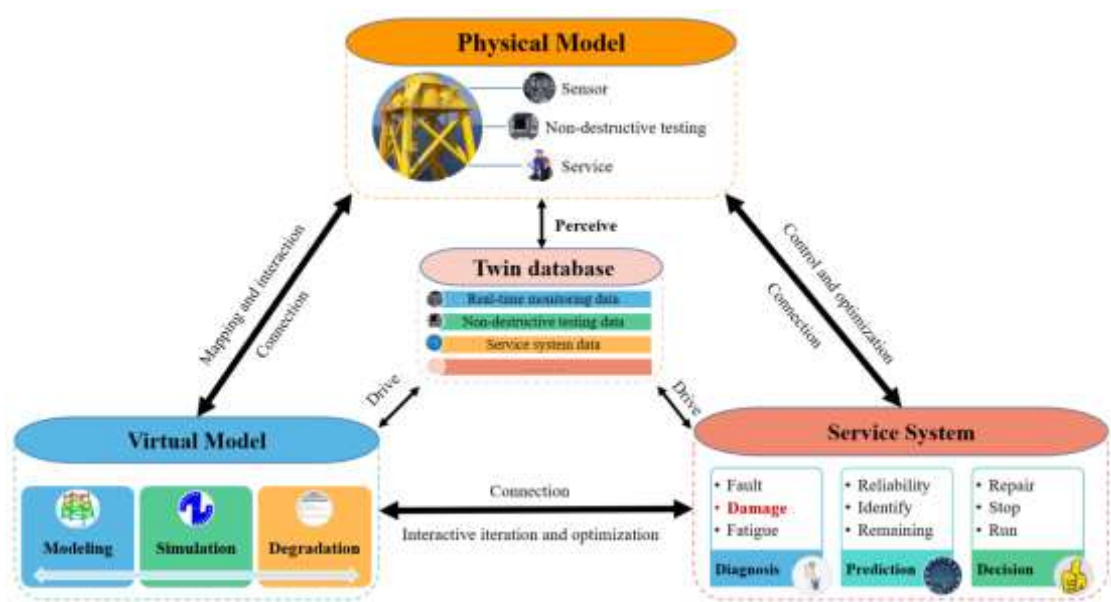

**Figure 1**. The schematic of the digital twin of the jacket platform.

The schematic of the digital twin (DT) of the offshore jacket platform has been proposed in [5], as shown in Fig.1. This DT has been established to enable various tasks such as damage identification, fault diagnosis, fatigue analysis, reduction of unnecessary maintenance activities, and formulation of optimal maintenance intervals. This DT consists of a physical model, a virtual model, a twin database and a service system. In the physical model, the sensor acquisition system must be installed and used to obtain the important parameters, such as load, dynamic response, damage, temperature and so on, and map them to the virtual model at the same time; then, the service system calls these parameters on demand to calculate the damage parameters or evaluate the information, and the result will feed back to the virtual model for model updating; finally, the updated high-precision virtual model performs an efficient prediction of the structural damage. Damage identification is a crucial component of health monitoring and an essential module in the service system of the DT framework, which can help to detect the presence of damage on a platform and assess its impact on the structural safety of the platform. Through damage identification, timely maintenance measures can be taken to reduce the incidence of failures, decrease maintenance costs and increase the reliability and safety of the platform. Therefore, the position and extent of the damage to the offshore jacket platform must be determined, making damage identification a crucial topic of research.

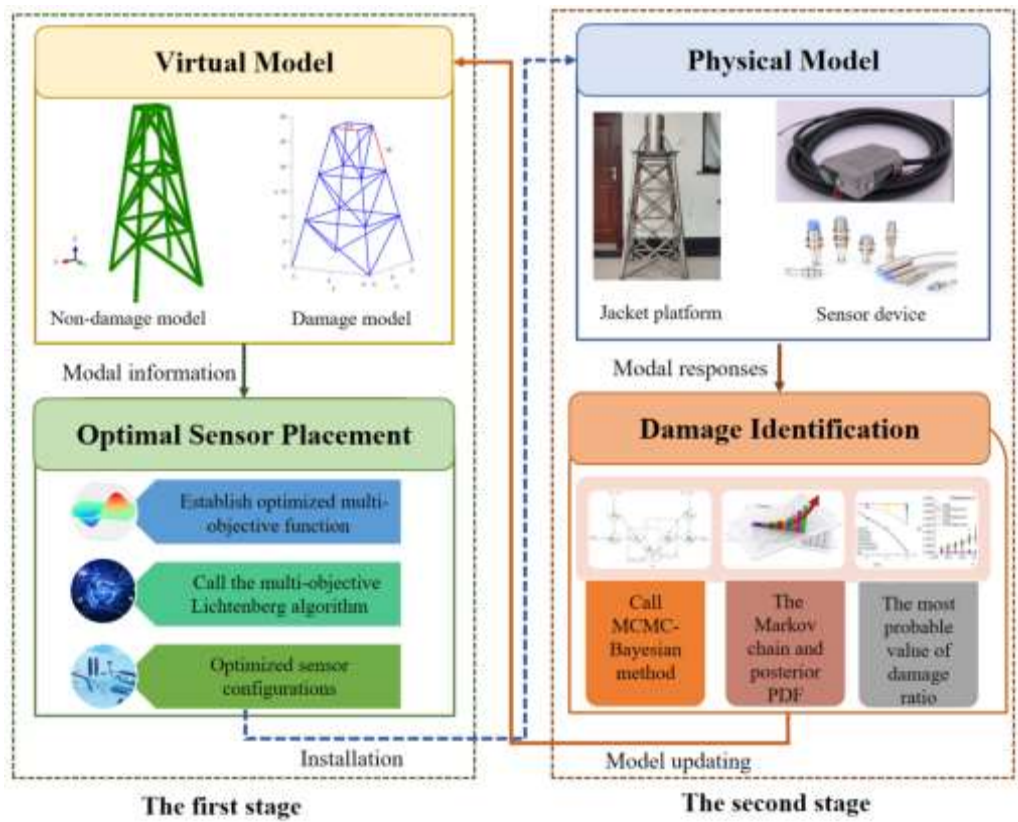

**Figure 2.** The framework of proposed damage identification.

Damage identification framework in Fig. 2 comprises physical and virtual models, an optimum sensor placement model, and a damage identification model, all of which are informed by the physical location of sensors. The optimal sensor placement model seeks to identify the optimum sensor configurations, including the optimal number and locations of sensors, while the damage detection model is used to ascertain the damage's location and severity. The two-step process of the suggested damage identification framework is described below. Before MOLA can be used to solve the Pareto front, the modal information needed to optimise the objective function must be calculated using the non-damage finite element model. The best possible sensing setups are then selected and implemented in the underlying physical model. The second step involves collecting data on the modal answers in the damaged state and using that information to feed into a damage identification model. As a next step, we use the Bayesian approach to compute the Markov chain and posterior probability distribution function (PDF) of the damage ratios, which provide insight into the spread of damage across the map. Simultaneously, the most likely values of the damage rates are fed back into the virtual model to update the model. The optimal sensor placement model and the damage identification model are explained below, along with their respective operational processes.

**2.2 Optimal sensor placement**

Obtaining the best possible sensor setups is depicted in a detailed flowchart in Fig. 3, which depicts the optimal sensor location model. The natural frequency and mode shapes, as well as other modal reactions, are computed in a non-damage finite element model and can be saved in a database for later use. After that, MOLA will be used to determine and resolve the goal functions.

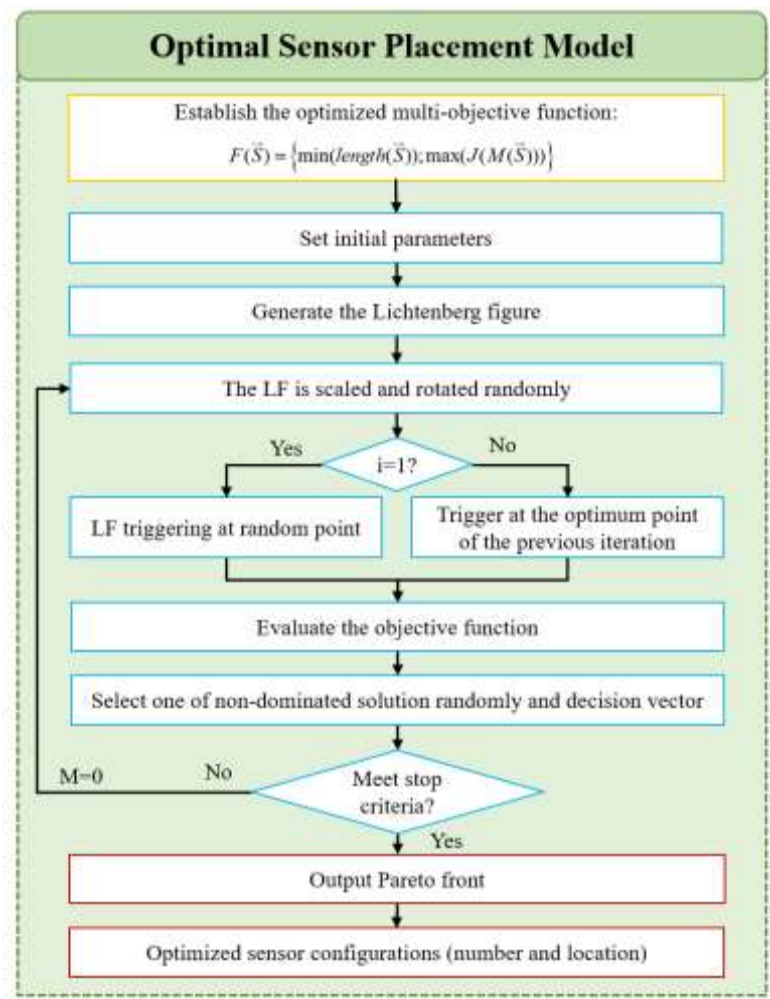

**Figure 3.** The flowchart of the optimal sensor placement in this paper.

In the optimal sensor placement model, four well-known modal metrics of the OSP are employed to construct the objective function to guarantee the independence and accuracy of modal information, which are Effective Independence (EFI), Kinetic Energy (KE), Eigenvalue Vector Product (EVP) and Information Entropy (IE). The details information of these modal metrics are as follows:

When it comes to big buildings, one of the most popular metrics is EFI [33], an efficient unbiased estimator that maximises the norm of the Fisher information matrix [31]. The covariance matrix of the prediction error is expressed as

$$E_D = \Phi^T \Psi \lambda^{-1} (\Phi \Psi)^{-1} = \Phi (\Phi^T \Phi)^{-1} \Phi^T \qquad (1)$$

Where $\Psi$ and $\lambda$ are the eigenvector and eigenvalue of structure, respectively. $\Phi$ is the mode shape matrix of the FEM. The larger the $E_D$, the greater contributions of the sensor position to the independence of the structural modes[47].

The KE metric [29] measures the dynamic contribution of each FEM element to each of the target mode shapes, as calculated by Equation (2):

$$KE_{in} = \Phi_{in} \sum_j M_{ij} \Phi_{jn} \omega_n^2 \qquad (2)$$

where $KE_{in}$ is the kinetic energy associated with the *i*-th degree of freedom in the *n*-th mode, $\Phi_{in}$ is the *i*-th coefficient in the *n*-th mode, $M_{ij}$ is the *i*-th row and *j*-th column element in the mass matrix, and $\Phi_{jn}$ is the *j*-th coefficient in the *n*-th mode [33].

The EVP determines the optimal measurement candidate by calculating the maximum of the product of the mode shapes at each of its places for the *N* modes being measured. The *i*-th EVP is calculated by [29]

$$EVP_i = \prod_{j=1}^{N} |\Phi_{ij}| \qquad (3)$$

The IE is a useful tool for determining which structural tests should be performed in order to lessen the impact of ambiguity [29]. The lower the information entropy, the higher the certainty of the system. The basic form of the IE is given by

$$IE(x) = -\sum_{i=1}^{N} P_i \log(P_i) \qquad (4)$$

Where $P_i$ is the corresponding probability of the *i*-th realization of discrete random variable and log(.) is the logarithm operator.

Another objective of the OSP is the number of sensors, which represents sensor deployment cost. As the number of sensors increases, the sensor installation cost and the processed data increase as well. Including the number of sensors as an objective not only explains the influence of the sensor number on the optimization criterion but also eliminates any sensor configurations that are dominated through the Pareto dominance relationship [29]. Moreover, combining the number of sensors and one of the four modal metrics as objective functions can keep a balance between sensor cost and the OSP quality.

To solve the conflicting objectives of maximizing the obtained modal information from the target system while minimizing the number of sensors, the MOLA [45] is first employed to solve the optimization problem of the offshore jacket platform. MOLA is a new menta-heuristic multi-objective optimization algorithm inspired by lightning storms and Lichtenberg Figures (LF), which has been successfully applied in crack detection and other areas. The details of the creation about the MOLA has been introduced in [45]. There are five key parameters that determines the construction and works of the MOLA, which are population (*pop*), the number of particle (*Np*), creation radius (*Rc*), stick coefficient (*S*), switching factor ( *M* ), refinement ( *ref* ) and iteration (*N$_{iter}$*). Among them, *Np* is used to create an LF based on the Diffusion Limited Aggregation theory, as shown in Fig. 4(a). *pop* is the number of points used to compute the objective functions and is described by black dots in Fig. 4(b). *Rc*, *Np* and *S* are three important parameters for the construction of LF. The former is associated with LF size, the last two controls the density of the LF created. *M* is used to change the LF in the optimizer input data, which can be worth zero, one or two. The *ref* is an input parameter that can be from 0 to 1. If *ref*=0, only the global one LF acts on the optimizer every iteration. The number of iterations (*N$_{iter}$*) is generally taking the value from 100 to 1000, which also defined as an initial configure parameter of the algorithm. Exploration and exploitation are well-represented by MOLA's results because the objective function evaluation points are fired at a Lichtenberg figure of varying sizes and orientations at each iteration [45].

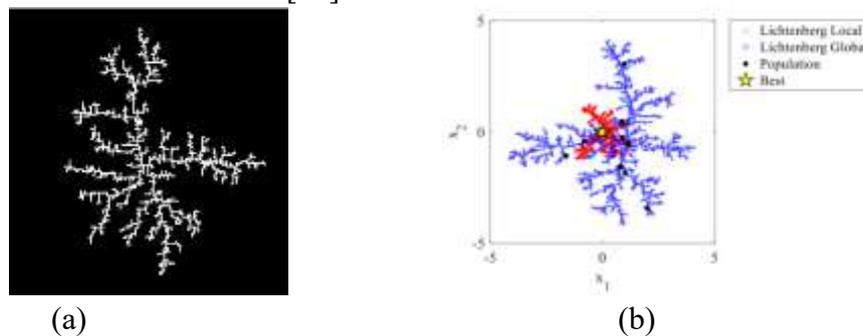

(a)          (b)

**Figure 4.** The Lichtenberg algorithm [48]. (a) bitmap Lichtenberg figure; (b) the population distribution.

All Pareto fronts are generated and compared by MOLA. The number and locations of sensors can be selected from all Pareto fronts. After determining the optimal sensor configurations, the sensors will be installed in the offshore jacket platform to collect the dynamic response.

**2.3 Damage identification model**

In damage identification model, the Markov Chain Monte Carlo (MCMC)-based Bayesian method [17,49] is used to calculate the damage ratio, and the detailed damage identification model is shown in Fig. 5, which includes one sample generation module generated the candidate samples by the proposed distribution and one Bayesian module quantified the uncertainty of the parameters. The detailed steps of the damage identification are introduced in the following.

In the first step, we treat the element damage ratio as the unknown parameters, and use the prior distribution to draw initial examples. The candidate samples are then created using the suggested probability distribution function and the initial values. (PDF).

Secondly, the FE model is used to calculate the natural frequencies and mode shapes for both the initial and candidate samples. The normalized errors are used in the proposed model, shown in the following:

$$\varepsilon_f = \frac{f_m - f_a}{f_a} \qquad (5)$$

$$\varepsilon_\varphi = \sqrt{1 - |\varphi_m^T \varphi_a|^2} \qquad (6)$$

where $f_m$ and $f_a$ respectively denote the measured and predicted natural frequencies; $\varphi_m$ and $\varphi_a$ respectively denote the measured and the predicted mode shapes.

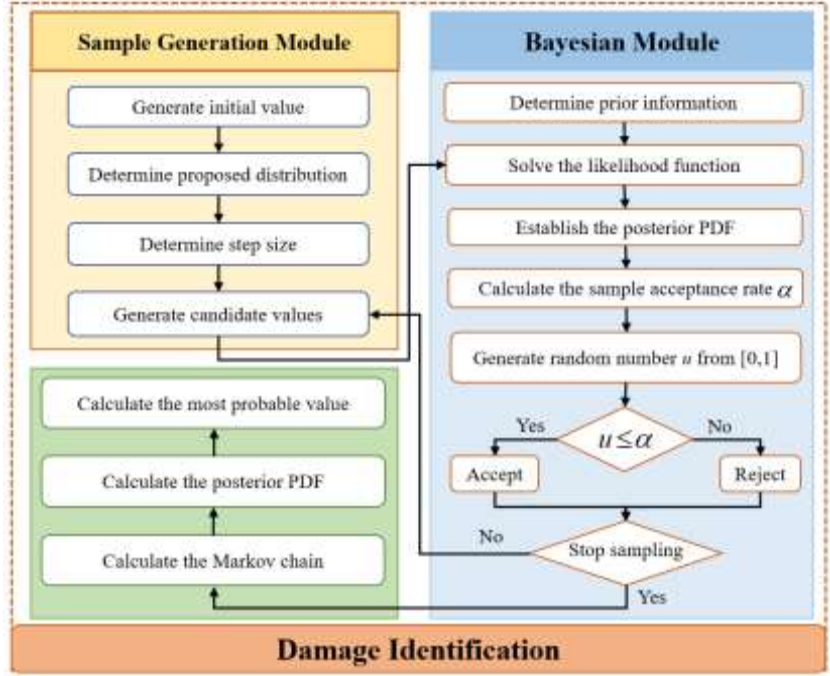

**Figure 5.** The flowchart of the damage identification

Thirdly, according to Bayesian theory [17,50,51] described in Equation (7), the posterior PDF is calculated using the observed modal data (from the physical model) and the calculated modal data (from the FE model). All modes are supposed to have independently distributed natural frequencies and mode shapes, as in the previous PDF. The posterior PDF of the damage ratios are calculated by Equation (8)-(9).

$$P(A/B) = \frac{P(A,B)}{P(B)} = \frac{P(B/A)P(A)}{P(B)} \propto P(B/A)P(A) \qquad (7)$$

where $A$ is the vector of the parameters being estimated, $P(A)$ is the prior distribution. The prior distribution is assumed to be a generalized and unbiased uniform distribution according to Bayesian hypothesis. $B$ is the measured information, and $P(B)$ is a normalizing constant. $P(B/A)$ is the likelihood function, and $P(A/B)$ is the posterior PDF and can be calculated by

$$P(A/B) \propto P(f/A)P(\varphi/A)P(A) = C \cdot \exp(-\frac{1}{2}J(A)) \qquad (8)$$

$$J(A) = \sum_{i=1}^{n}[(\frac{f_m^i - f_a^i}{f_m^i})^2 + (1 - |\varphi_m^T \varphi_a|^2)] \qquad (9)$$

where $n$ is the number of measured modes and $C$ is a normalizing constant.

Lastly, the Metropolis-Hastings (MH) method [52] is adopted to sample and approximate the posterior PDF of the damage ratio and the Markov chain is used calculated to evaluate the most probable value of the damage ratio based on the posterior PDF. Afterwards, the damage location and severity can be identified and fed back into the DT virtual model for model updating.

## 3. The OSP of the offshore jacket platform

To verify the efficacy of the optimal sensor location model, a numerical case study is implemented in this subsection. Modal details and background on the offshore garment platform are presented in Section 3.1. The optimum sensor placement is displayed in Section 3.2.

### 3.1 Description of the offshore jacket platform

Based on the work of Liu et al. [53], a three-dimensional finite element model (FEM) of an offshore jacket structure is developed using MATLAB codes (see Fig. 6). The jacket platform is represented by a model based on the Euler-Bernoulli beam (B31) theory, which comprises of 36 pipe elements and 16 nodes. The model has 72 DOF (depending on how you count), with each component having 6 DOF. The offshore jacket platform has four levels and measures in at a total height of 9.14 meters, with side lengths of 10.97 meters, 8.53 meters, 6.10 meters, and 3.66 metres from lowest to highest level. The outer diameter of each pipe part is 17.8 cm and the inner diameter is 16.02 cm. The material of the jacket structure is steel (i.e., elastic stiffness is 210 GPa, mass density is 7,850 kg/m$^3$, and Poisson's ratio is 0.3).

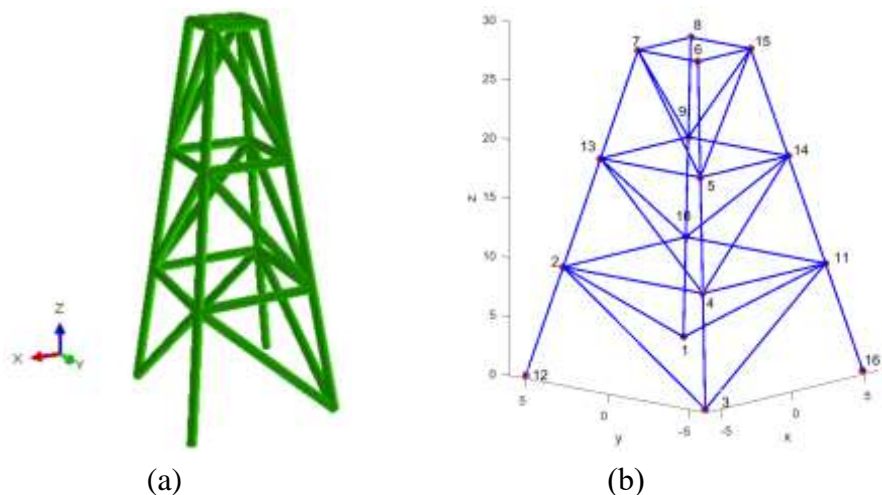

(a)          (b)

**Figure 6.** The model of the offshore jacket platform. (a) FE model; (b) the number of nodes.

The objective functions in this research are built using the first six natural harmonics and mode shapes. The modal information is calculated under non-damage state, as presented in Table 1 and Fig. 7.

**Table 1.** The natural frequency of the offshore jacket platform

| Modes | 1 | 2 | 3 | 4 | 5 | 6 |
|---|---|---|---|---|---|---|
| Natural frequency | 6.9989 | 9.5169 | 9.7781 | 14.4633 | 16.8264 | 18.2163 |

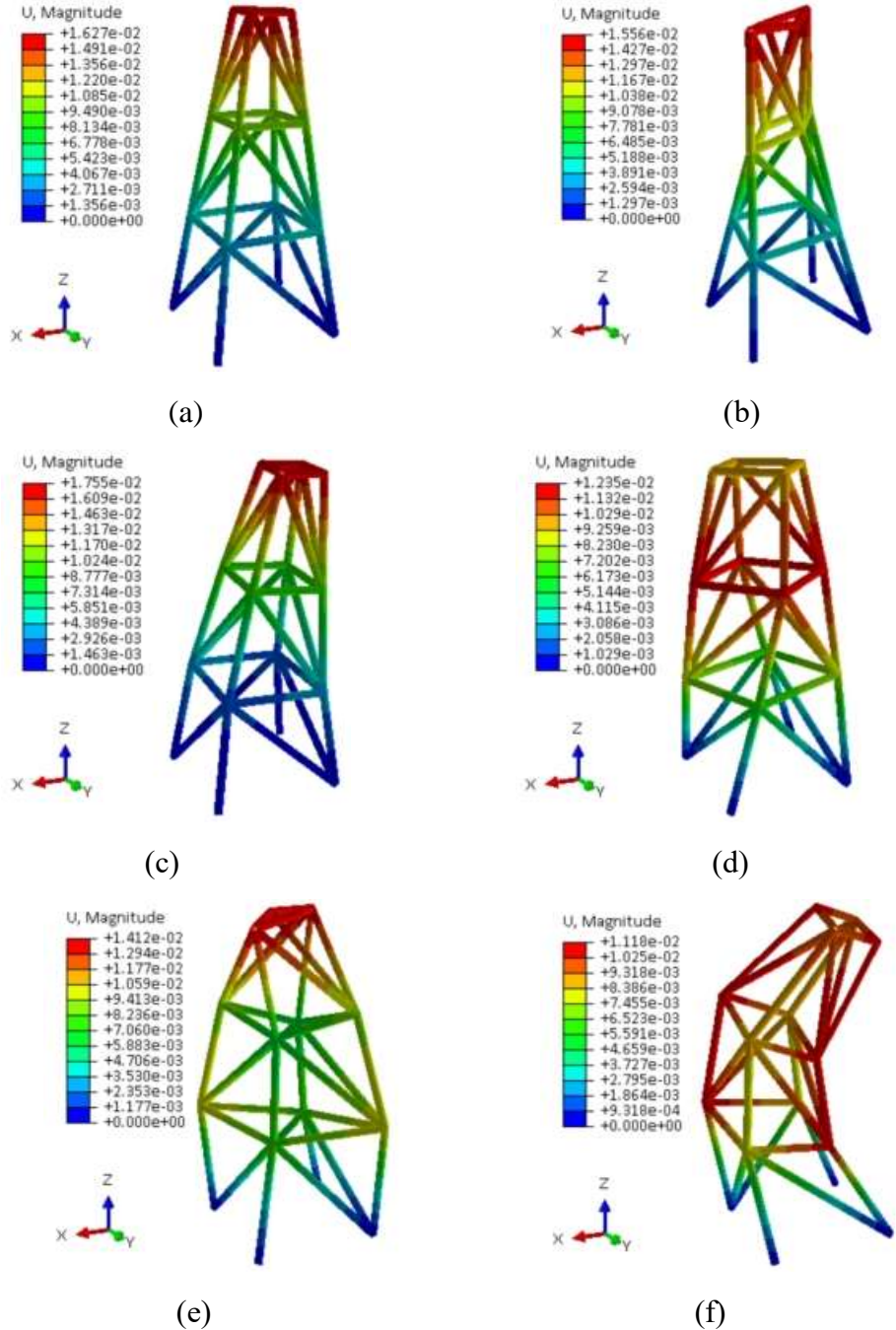

(a)  (b)

(c)  (d)

(e)  (f)

**Figure 7.** The first six model shapes of the offshore jacket platform.

### 3.2 Optimal sensor placement

In this paper, for the optimization of sensor placement, the MOLA [29,45,48] is adopted to determine the optimal sensor configurations of the offshore jacket platform. Initially, 12 candidate sensor locations with No. 2,4,5,6,7,8,9,10,11,13,14,15 are selected and shown in Fig. 6. Since these positions are the intersections of the structural system of the offshore jacket platform, where the stress and deformation are most critical and likely to occur. By placing sensors at these locations, we can more accurately monitor the stress and deformation of these critical areas, detect potential issues in a timely manner, and take corresponding measures to repair or reinforce them, ensuring the safe operation of the platform. The number of sensors and one of four well-known optimization criterions (KE, EVP, IE and EFI) are considered as the objective functions. The optimization objective is described as

$$F(\vec{S}) = \{\min(length(\vec{S})); \max(J(M(\vec{S})))\} \quad (10)$$

Where :

$$\vec{S} = \vec{A} * \vec{B^T}$$
$$\vec{A} = \{1, 2, ...n\}$$
$$\vec{B} = \{a_1, a_2, ...a_n\}$$

A vector $\vec{S}$ is generated for each iteration, which is used to evaluate two objective functions: sum of the number of sensors and in the objective function related to the metric $J(M)$. $\vec{A}$ is the vector of candidate sensors, at the same time, the MOLA generates another binary vector of the same length $\vec{B}$ to generate the selected sensor vector $\vec{S}$.

The parameters of MOLA are set according to Table 2 and *v*-shaped is employed as the transfer function [54,55].

Table 2. The value of control parameters of MOLA.

| Parameter | pop | Np  | Rc  | S | M | ref | $N_{iter}$ |
|-----------|-----|-----|-----|---|---|-----|------------|
| Value     | 100 | 900 | 100 | 1 | 0 | 0.5 | 50         |

The Pareto dominance relationship is then used to select the optimal sensor placement options in the objective space (shown in Fig.8). The figures show that anywhere from one to all twelve of the possible sensors can be optimally placed according to the four factors. For the KE, IE, and EFI criteria, we find that the optimal number of sensors grows linearly with the amount of information gathered in the jacket structure. As the number of sensors increases, however, the EVP criterion forms a convex Pareto front. As can be seen in Fig.8(d), when fewer than five sensors are used, the EVP criterion value responds strongly to the addition of sensors, but with more than eight sensors, there is no notable improvement. Additionally, as more instruments are added to the offshore jacket platform, a higher quality of data is collected.

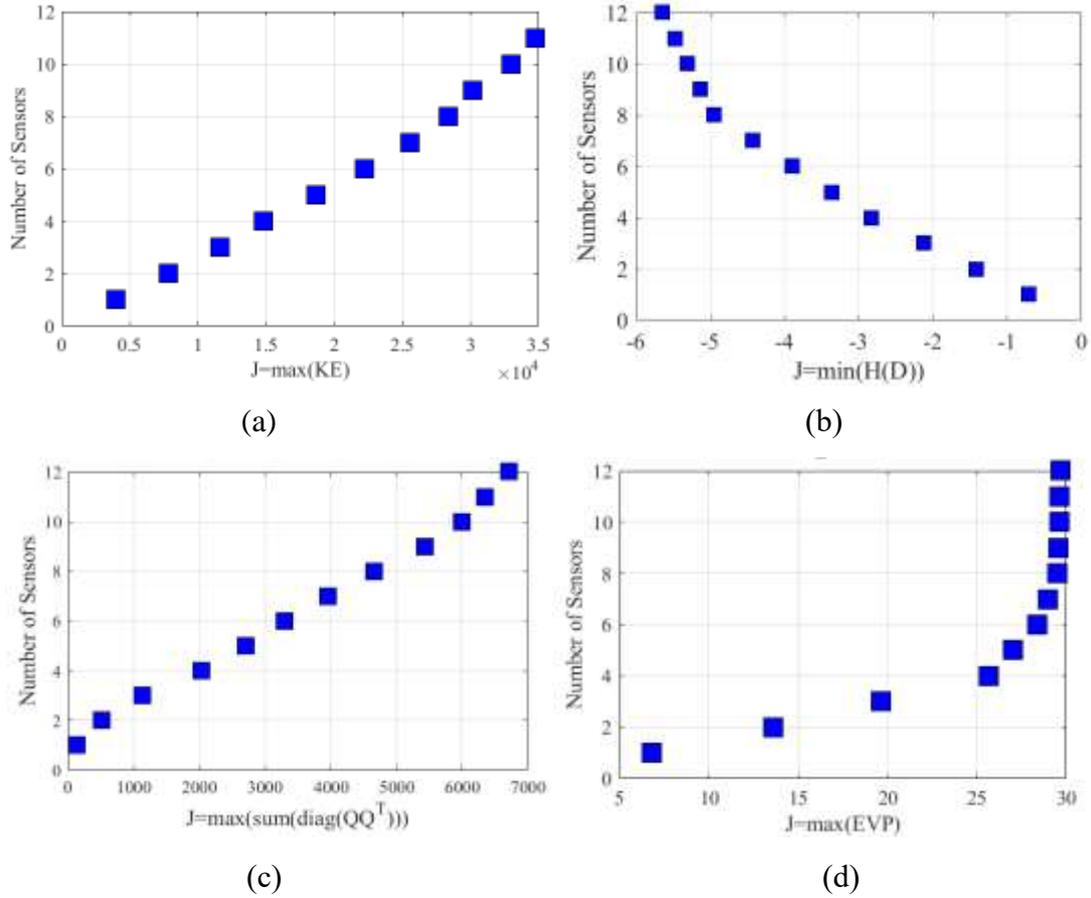

**Figure 8.** The relationship of modal criterion and the number of sensors in the Pareto front. (a) KE; (b) IE; (c) EFI; (d) EVP.

Moreover, the Hypervolume metric is used to compare the different Pareto front families with each other [29]. A higher Hypervolume indicates better convergence and coverage of the Pareto front. The average results after 10 independent runs of all indicators are shown in Fig. 9, and it is obviously that EVP criterion has a higher Hypervolume.

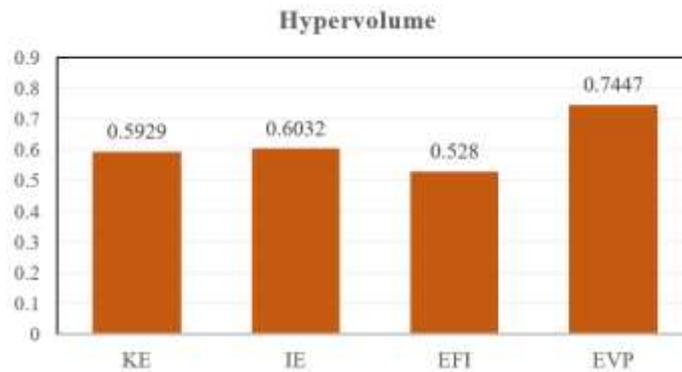

**Figure 9**. The Hypervolume values

Although four criterions of optimal sensor placement can find solutions, the best criterion is EVP as it presented the highest Hypervolume and indicates that criterion has greatest convergence and coverage. Consequently, EVP is applied to optimal sensor placement.

In order to show the superiority of the MOLA, NSGA-II is also employed to compare with MOLA, which is a widely used multi-objective optimization problem in various real word applications. Using a particular form of crossover and mutation, NSGA-II produces offspring from which the next generation is chosen using nondominated-sorting and crowding distance comparison [56]. Take the EVP metric as an example, the key parameters of NSGA-II algorithm in the offshore jacket platform example are set as follows. The population size is 200 and the maximum number of generations is 100. The crossover function is scattered, and the probability of crossover is 0.8. The mutation function is Gaussian, and the probability of mutation is 0.6, the mutation strength is 0.1. The Pareto front calculated by NSGA-II is shown in Fig. 10. It can be seen that the result is similar with that obtained by MOLA. The EVP criterion is not changed as the number of sensors increased beyond 8 sensors. Therefore, 8 sensors can be chosen as the minimum number.

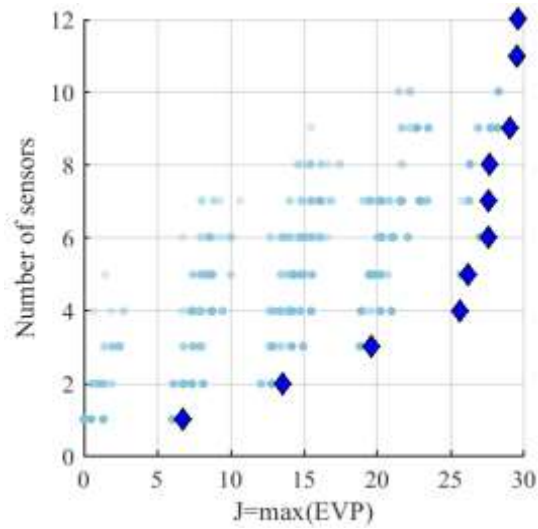

**Figure 10**. The Pareto front with EVP criterion.

Table 3 presents the sensor positions selected by MOLA and NSGA-II for eight sensors. It can be observed that the top four sensor positions are common to both algorithms. However, there is a difference in the selection of the remaining sensor positions. MOLA tends to place sensors on the second layer, while NSGA-II focuses on the third layer. This indicates that MOLA is more inclined to maximize the response amplitude by placing more sensors on the top of the offshore jacket platform. In terms of the target EVP criterion J(M), MOLA outperforms NSGA-II. Therefore, MOLA is considered superior to NSGA-II and the sensor configurations obtained by MOLA will be used for damage identification.

Table 3. The sensor configurations with EVP metric

| Algorithm | Sensor positions | J(M) |
| --- | --- | --- |
| MOLA | 4,5,6,7,8,9,14,15 | 29.49 |
| NSGA-II | 2,4,6,7,8,10,14,15 | 19.63 |

## 4. Damage identification using MCMC-Bayesian method

### 4.1 Damage identification

The proportion of stiffness lost is used as a measure of damage. The change of the $n$ element stiffness can be expressed by Equation (11)

$$\Delta K = (1-\alpha_n)K_n \quad (11)$$

where $\alpha_n$ is the changes in the stiffness, called as the damage ratio (0~1), $K_n$ is the initial stiffness of the $n$-th element.

In this work, to quantify the noise effect, random errors were added to the natural frequencies and mode shapes. The measurement errors are simulated as follows [57]:

$$f_p^{(n)} = f_p(1+y_f \times \varepsilon) \quad (12)$$

$$\varphi_{p,q}^{(n)} = \varphi_{p,q}(1+y_\varphi \times \eta) \quad (13)$$

where $f$ and $\varphi$ are the natural frequency and mode shape, respectively; $y$ is the random number with zero mean value and standard deviation of 1; superscript $n$ denotes noisy parameters; $\eta$ and $\varepsilon$ represent the noise level in the mode shape and natural frequency, respectively.

**Table 4.** The simulated damage cases

| Damage cases | Description |
|---|---|
| D1 | E3=0.8 |
| D2 | E3=0.6, E6=0.4, E9=0.2 |

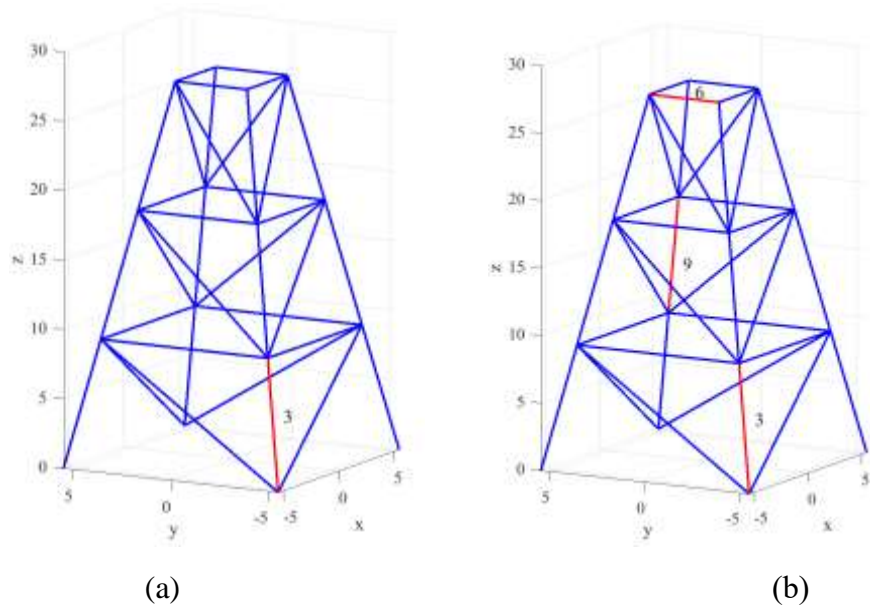

(a)            (b)

**Figure 11**. The damage scenarios. (a) one damage (D1); (b) three damage (D2).

Damage is replicated by lowering the part elements' elasticity modulus. Table 4 details two damage scenarios, and Figure 11 identifies the components of the jacket structure that were compromised. Using the FE model, a modal analysis is computed, with the native frequency and mode shapes distorted by 1% noise. Modal information from the first six phases is used for damage diagnosis.

In the damage identification process, the MCMC-based Bayesian method is employed for damage identification. Several parameters of MCMC-based Bayesian method are set as follows. The MH algorithm is set to terminate after reaching 15,000 iterations. The first 5,000 iterations are prepared for the non-adapting period while the remaining 10,000 iterations are used to form stable Markov chain. The most probable value of the damage ratio is obtained as the average of the 10,000 samples.

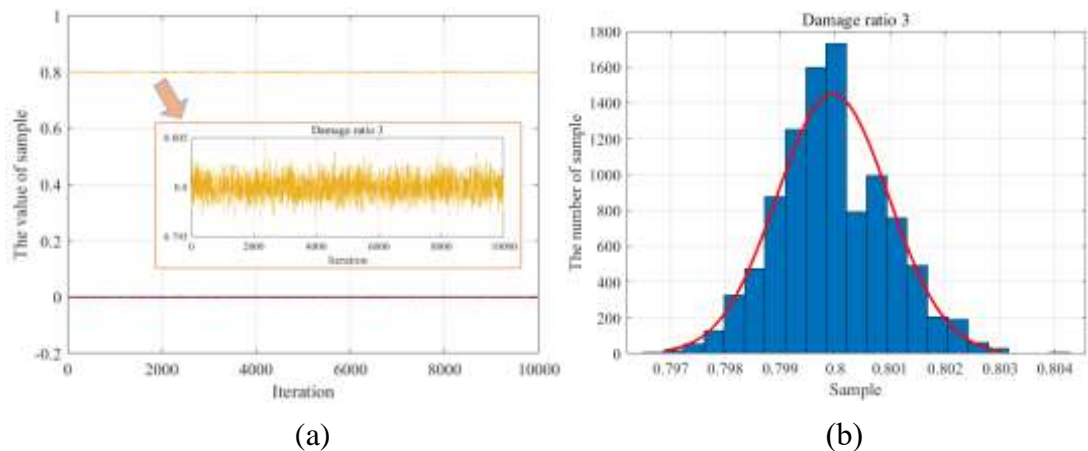

(a) (b)

Figure 12. The damage identification result for one damage. (a) the Markov chain of damage ratio; (b) the posterior PDF of damage ratio 3.

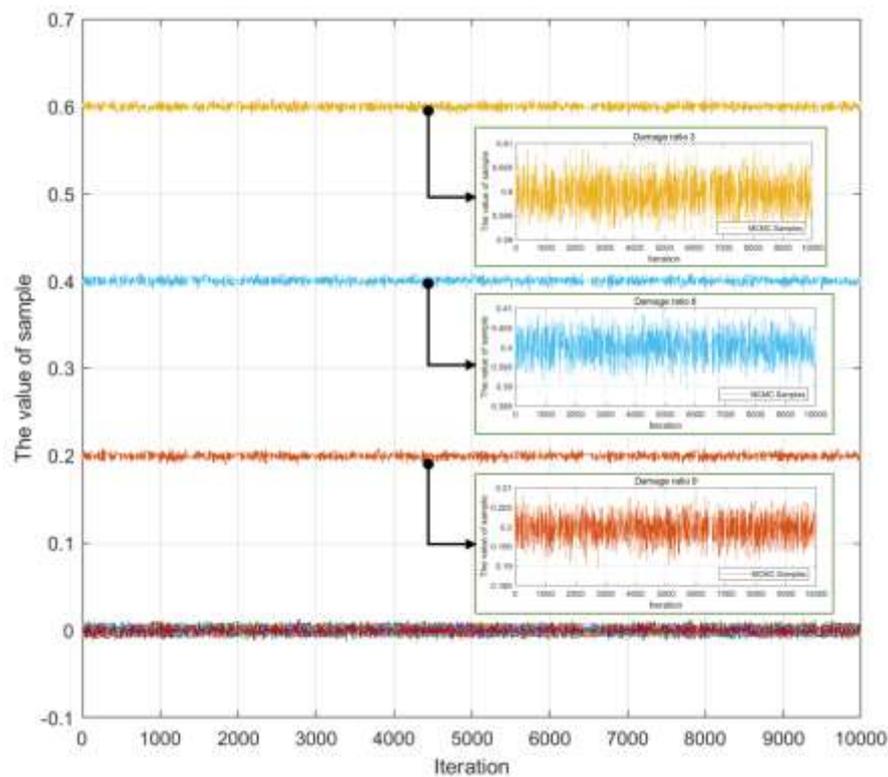

Figure 13. The Markov chain of damage ratio with three damage scenarios (D2).

For damage scenario (D1) with one damage location, the Markov chain of damage ratio is shown in Fig. 12(a). It can be observed that all samples keep fluctuating around the stable value in the sampling process. The damage ratio of element 3 (damage ratio 3) converges to 0.8, while another damage ratios converge to 0. As shown in Fig. 12, the kernel density estimation is then used to compute the Bayesian marginal PDF for the damage ratio 3.(b). The posterior marginal PDF of the damage ratio 3 fits neatly into the Gaussian distribution, as can be seen. The most probable value of damage ratio 3 is 0.8 and the errors of the damage identification are close to zero. Consequently, the damage ratio of element 3 is identified successfully.

For damage scenario (D2) with three damage locations, the Markov chain is calculated and shown in Fig. 13. It can be seen that the damage ratio 3 converges to 0.6; the damage ratio 6 converges to 0.4; the damage ratio 9 converges to 0.2, while the other damage ratios converge to 0. The above results indicate that there are three different degrees damage in the offshore jacket platform. The posterior PDF of damage ratios 3, 6 and 9 are calculated and shown in Fig. 14. At the same time, the 95% confidence interval can be observed in the Fig.15. The results demonstrate that the Bayesian method has a great potential on the damage identification, as it allows for quantification of the uncertainty related to the identified variables.

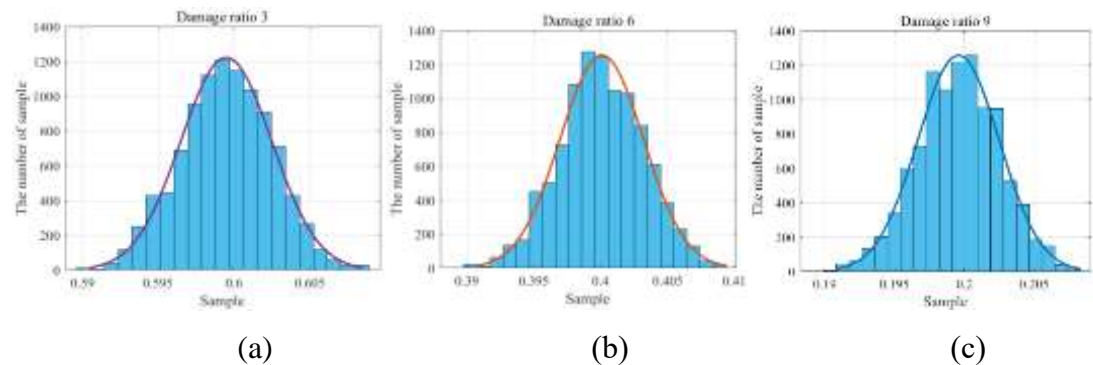

(a)                      (b)                      (c)

**Figure 14.** The posterior PDF of damage ratio. (a) damage ratio 3; (b) damage ratio 6; (c) damage ratio 9.

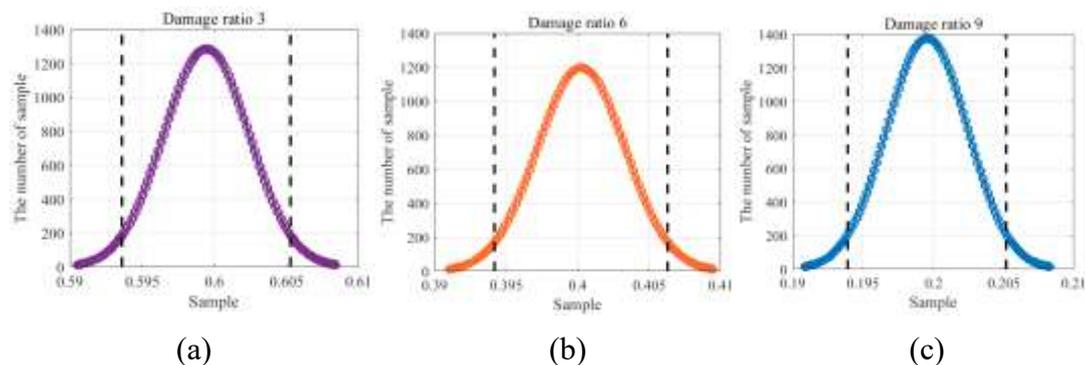

(a)                      (b)                      (c)

**Figure 15.** The posterior PDF of damage ratio with 95% confidence interval.

The most probable values and the identified error of the damage ratios are both calculated and shown in Table 5. To demonstrate the superiority of OSP, the modal information collected by all candidate sensor locations are also used for damage identification, and the results are shown in Table 5. The identification results are very close in both situations, but the calculation efficiency is higher and the sensor cost is reduced for the OSP. All the results demonstrate that the proposed the damage identification framework not only identify single damage situation but also handles multiple damage situation and is effective for the platform damage identification.

Table 5. The identification result of damage ratio.

| Variables | Optimal sensor layout (error) | Candidate sensor location (error) |
|---|---|---|
| Damage ratio 3 | 0.599 (0.17%) | 0.599 (0.17%) |
| Damage ratio 6 | 0.401 (0.25%) | 0.400 (0%) |
| Damage ratio 9 | 0.198 (1.00%) | 0.199 (0.50%) |
| Number of sensors | 8 | 12 |

### 4.2 The influence of measurement noise

Modal information measurements are highly sensitive to noise. The suggested damage identification framework requires further study into the impact of noise.

As an illustration, let's say we have two damage conditions, and we decide to make the damage ratios for elements 3 and 9 0.8 and 0.5, respectively. Modal properties are calculated numerically, and then white Gaussian noise is added to them to mimic the impacts of measurement uncertainty in accordance with Equations (12) and (13). (13). We introduce random variation to the modal data at 1%, 5%, 10%, and 15% values. Five iterations of the simulation—the functional equal of five experiments—are run with noise added in order to collect enough information for damage identification. See the impact of noise on the suggested damage identification framework by comparing the natural frequencies with varying noise levels (Fig. 16).

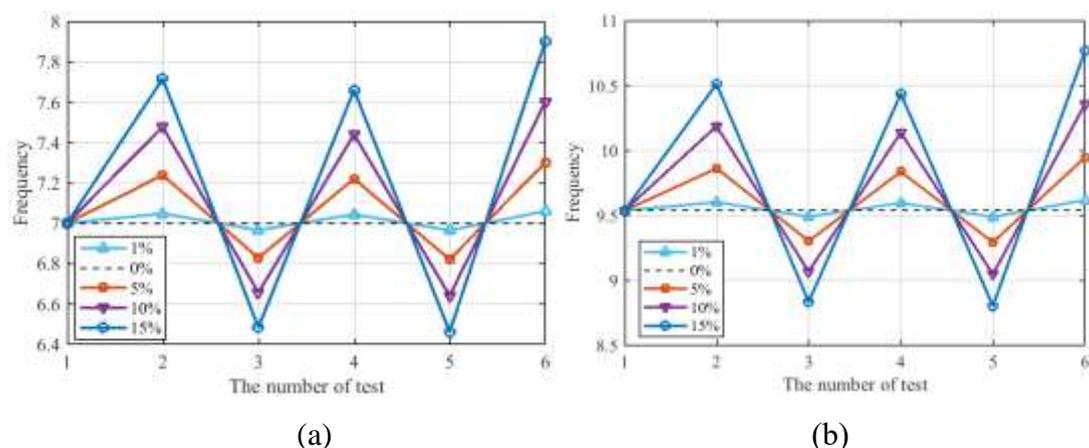

(a)          (b)

**Figure 16.** The natural frequency with different levels of noise. (a) the first order; (b) the third order.

Then, the MCMC-based Bayesian method is employed to calculate the damage ratio with different noise levels. The corresponding posterior marginal PDFs are presented in Fig. 17, which reveals that all PDFs follow Gaussian distributions. However, when the noise level increases, the peak of the posterior PDFs deviates from the true values.

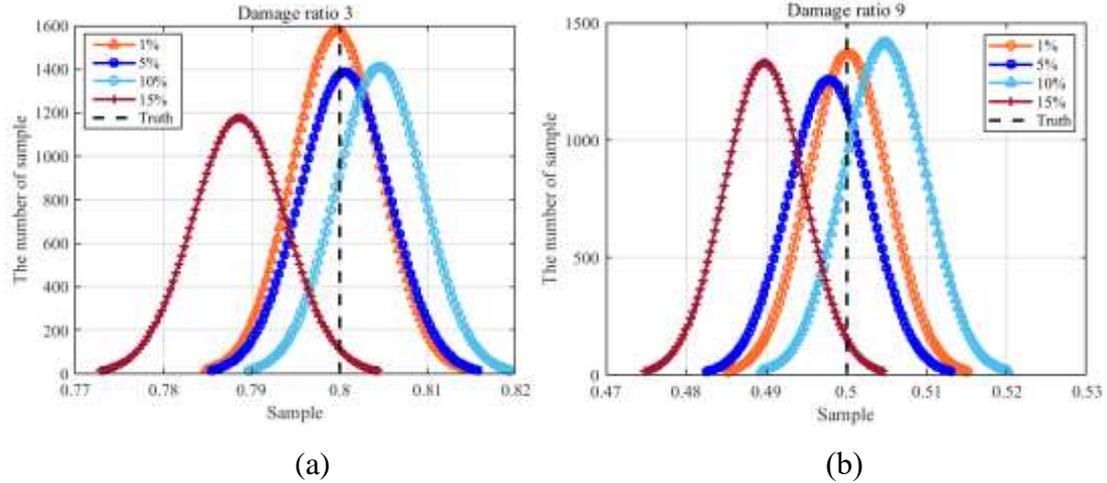

(a)                  (b)

**Figure 17**. The posterior marginal PDF with different noise levels. (a) damage ratio 3; (b) damage ratio 9.

To be more specific, the identification error of damage ratios under different noise levels is shown in Fig. 18. As expected, the errors gradually increase with higher levels of measurement noise. The maximum error of 2% is observed for damage ratio 9 when the noise level is set at 15%, which is still acceptable for the damage identification process. These results demonstrate the robustness and applicability of the proposed damage identification framework for offshore jacket platforms.

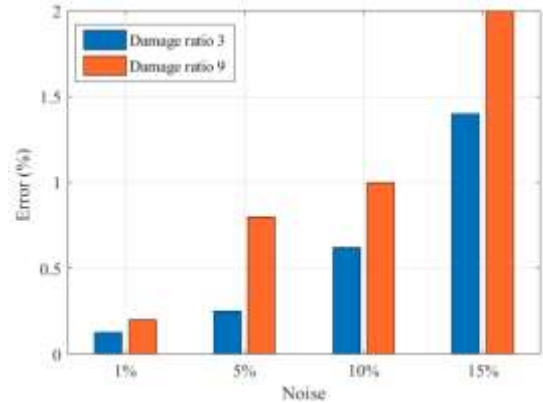

**Figure 18**. The error of the damage identification with different noise levels.

## 5. Conclusions

To ensure precise health monitoring of the offshore jacket platform, a novel damage identification framework is created in this research. The framework combined the optimal sensor placement model with the damage detection model to pinpoint the exact location and severity of any damage. MOLA was used to find a happy medium between the quantity and quality of OSP sensors, thereby solving the bi-objective issue inherent in the optimal sensor placement model. To efficiently and correctly compute the damage ratio and quantify uncertainties of model parameters, the damage identification model used the MCMC-based Bayesian method. Above 97% accuracy in damage identification within a noise level of 15% and reduced sensor cost were proven in the case study, proving the efficacy of the proposed framework. Future work includes incorporating the damage detection model into the DT framework of offshore jacket platforms to create a comprehensive health monitoring system based on the identified damage ratio used for model updating in the virtual model. In order to create the DT framework for health monitoring of offshore jacket platforms, the proposed damage identification framework offers technical assistance.

## Acknowledgements

This research is supported by the NSFC (51979261) and Natural Science Foundation of Hebei Province of China (A2020202017), Youth Foundation of Hebei Education Department (QN2020211).